\documentclass[final,3p,times]{elsarticle}

\usepackage{amssymb}

\usepackage{amssymb}
\usepackage[]{amsmath}
\usepackage{graphics}
\usepackage{amssymb}
\usepackage[]{amsmath}
\usepackage{amsthm}
\usepackage{setspace}
\usepackage{epsfig}
\usepackage{subfigure}
\usepackage{cases}
\usepackage{graphicx}

\biboptions{numbers,sort&compress}

\usepackage{amsmath}
\usepackage{times}
\usepackage{anysize}
\marginsize{2.5cm}{2.5cm}{1cm}{2cm}

\selectfont

\begin{document}

\begin{frontmatter}


\title{General high-order rogue waves of the (1+1)-dimensional Yajima-Oikawa system}

\author{Junchao Chen \fnref{label11,label1}}
\author{Yong Chen \fnref{label1} }
\ead{ychen@sei.ecnu.edu.cn}
\author{Bao-Feng Feng \fnref{label2} \corref{cor1}}
{\ead{baofeng.feng@utrgv.edu}}
\author{Ken-ichi Maruno \fnref{label3}}
\ead{kmaruno@waseda.jp}
\author{Yasuhiro Ohta \fnref{label4}}
\ead{ohta@math.kobe-u.ac.jp}


\cortext[cor1]{Corresponding author.}
\address[label11]{Department of Mathematics, Lishui University, Lishui, 323000, People¡¯s Republic of China}
\address[label1]{ Shanghai Key Laboratory of Trustworthy Computing, East China Normal University, Shanghai, 200062, People's Republic of China}
\address[label2]{Department of Mathematics, The University of Texas Rio Grande Valley, Edinburg,
TX 78539, USA}
\address[label3]{Department of Applied Mathematics, School of Fundamental Science and Engineering,
Waseda University, 3-4-1 Okubo, Shinjuku-ku, Tokyo 169-8555, Japan}

\address[label4]{Department of Mathematics, Kobe University, Rokko, Kobe 657-8501, Japan}

\begin{abstract}
General high-order rogue wave solutions for the (1+1)-dimensional Yajima-Oikawa (YO) system are derived by using Hirota's bilinear method and  the KP-hierarchy reduction technique.
These rogue wave solutions are presented in terms of determinants in which the elements are algebraic expressions.
The dynamics of first and higher-order rogue wave are investigated in details for different values of the free parameters.
It is shown that the fundamental (first-order) rogue waves can be classified into three different patterns: bright, intermediate and dark ones.
The high-order rogue waves correspond to the superposition of fundamental rogue waves.
Especially, compared with the nonlinear Sch\"{o}dinger equation,
there exists an essential parameter $\alpha$ to control the pattern of rogue wave for both first- and high-order  rogue waves since the YO system does not possess the Galilean invariance.

\end{abstract}

\end{frontmatter}


\section{Introduction}
Rogue waves, which are initially used for the vivid description of the spontaneous and monstrous ocean surface waves \cite{kharif2009rogue},
have recently attracted considerable attention on both experimentally and theoretically.
Rogue waves have been observed in a variety of different fields,
including optical systems \cite{solli2007optical,hohmann2010freak,montina2009non}, Bose--Einstein condensates \cite{bludov2009matter,bludov2010vector},
superfluids \cite{ganshin2008observation}, plasma \cite{moslem2011langmuir,bailung2011observation}, capillary waves \cite{shats2010capillary} and even in finance \cite{yan2011vector}.
Compared with the stable solitons, rogue waves are the localized structures with the instability and unpredictability \cite{akhmediev2009waves,akhmediev2009extreme}.
A typical model for characterizing the rogue wave is the
celebrated nonlinear Schr\"{o}dinger (NLS) equation.
The most fundamental rogue wave of the NLS equation is described by Peregrine soliton \cite{peregrine1983water},
which is first-order rogue wave and expressed in a simple rational form including the polynomials up to second order.
This rational solution has localized behavior in both space and time, and its maximum amplitude attains three times the constant background.
The Peregrine soliton can be obtained from a breather solution
when the period is taken to infinity.
More recently, significant progress on higher order rogue waves has been achieved \cite{akhmediev2009rogue,kedziora2012second,ankiewicz2011rogue,kedziora2011circular,dubard2010multi,dubard2011multi,
gaillard2011families,guo2012nonlinear,ohta2012general,ankiewicz2010rogue,he2010generating,mu2016dynamic,ling2016multisoliton,wang2017dynamics,chen2015rational,
bludov2009rogue,ankiewicz2010discrete,ohta2014general,yan2010nonautonomous,yan2010threedimensional,yan2010optical,wen2015generalized,yan2015twodimensional,yang2015rogue,wen2016dynamics,wen2017higherorder}
since a few special higher order rogue waves from first to fourth order were provided
theoretically by Akhmediev et al.\cite{akhmediev2009rogue} via the Darboux transformation method.
The higher order rogue waves were also excited experimentally in a water wave tank \cite{chabchoub2013observation,chabchoub2012observation}, which guarantees that such nonlinear
complicated waves are meaningful physically.
In fact, higher-order rogue waves can be
treated as the nonlinear superposition of fundamental rogue wave
and they are usually expressed in terms of complicated higher-order rational polynomials.
These higher-order waves were also localized in both coordinates and could exhibit higher peak amplitudes or multiple intensity peaks.

Another major development of importance is
the study of rogue waves in multicomponent coupled systems,
as a lot of complex physical systems usually contain interacting wave components with different
modes and frequencies\cite{ling2012highorder,guo2011roguewave,baronio2012solutions,zhao2013roguewave,baronio2013solutions,zhao2016high,ling2016darboux,
mu2015dynamics,zhai2013multirogue,wang2015roguewave,zhang2017solitons,ohta2012rogue,ohta2013dynamics,mu2017solitons,wen2015modulational,wen2016higherorder}.
As stated in Ref.\cite{ling2012highorder}, the
cross-phase modulation term in coupled systems leads to the varying instability regime
characters.
Due to the additional degrees of freedom, there exist more abundant pattern
structures and dynamics characters for rogue waves in coupled systems.
For instance, in the scalar NLS equation, because the existence of Galilean invariance, the velocity
of the background field does not influence the pattern of rogue waves. However, for the coupled
NLS system, the relative velocity between different component
fields has real physical effects, and cannot be removed by
any trivial transformation.
This fact bring some novel patterns for rogue waves
such as dark rogue waves \cite{guo2011roguewave}, the interaction between rogue waves
and other nonlinear waves \cite{guo2011roguewave,baronio2012solutions}, a four-petaled flower
structure \cite{zhao2013roguewave} and so on.
In particular, those more various higher order rogue waves in coupled nonlinear models enrich
the realization and understanding of the mechanisms underlying the complex dynamics of rogue waves.

Among coupled wave dynamics systems, the long-wave-short-wave resonance interaction (LSRI) is a fascinating physical process in which a resonant interaction takes place between a weakly dispersive long-wave (LW) and a short-wave (SW) packet when the phase velocity of the former exactly or almost matches the group velocity of the latter.
The theoretical investigation of this LSRI was first done by
Zakharov \cite{zakharov1972collapse} on Langmuir waves in plasma.
In the case of long wave propagating in one direction, the general Zakharov
system was reduced to the one-dimensional (1D) Yajima-Oikawa (YO) system\cite{yajima1976formation}.
This phenomenon has been predicted in diverse areas such as plasma physics \cite{zakharov1972collapse,yajima1976formation}, hydrodynamics \cite{grimshaw1977modulation,djordjevic1977two,ma1979some} and nonlinear optics \cite{kivshar1992stable,chowdhury2008long}.
For instance, this resonance interaction can occur
between the long gravity wave and the capillary-gravity one \cite{djordjevic1977two}, and between long and short internal waves \cite{grimshaw1977modulation} in hydrodynamics.
In a second-order nonlinear negative refractive index medium,
it can be achieved when the short wave lies on the negative index branch while
the long wave resides in the positive index branch \cite{chowdhury2008long}.
The (1+1) dimensional model equation, which is known as 1D YO system or LSRI system, can be written in a dimensionless form
\begin{eqnarray}
\label{yoequation01} && {\rm i}S_t - S_{xx} +SL=0,\\
\label{yoequation02} && L_t=-4(|S|^2)_x,
\end{eqnarray}
where $S$ and $L$ represent the short wave and long wave component, respectively.
The 1D YO system was shown to be integrable with a Lax pair, and was solved by
the inverse scattering transform method \cite{yajima1976formation}.
It admits both bright and dark soliton solutions \cite{ma1979some,ma1978complete}.
In Refs.\cite{cheng1992constraints,loris1997bilinear}, it is shown that the 1D YO system can be
derived from the so-called $k$-constrained KP hierarchy with $k=2$ while the NLS equation
with $k=1$.
Very recently, the first-order rogue wave solutions
to the 1D YO system have been derived by using the
Hirota¡¯s bilinear method \cite{wing2013rogue} and Darboux transformation \cite{chen2014dark,chen2014darboux}.
These vector parametric solutions indicate interesting structures that the long wave always keeps a single hump structure, whereas the short-wave field can be manifested as bright, intermediate and dark rogue wave.
Nevertheless, as far as we know, there is no report about high-order rogue wave solutions for
the 1D YO system.
Therefore, it is the objective of present paper to study high-order rogue wave solutions
of the 1D YO system (\ref{yoequation01})--(\ref{yoequation02}) by using
the bilinear method in the framework of KP-hierarchy reduction. As will be shown in the subsequent section,
a general rogue wave solutions in the form of Gram determinant is derived based on Hirota's bilinear method and the KP-hierarchy reduction technique. This determinant solution can generate rogue waves of any order without singularity.

The remainder of this paper is organized as follows.
In Section 2, we start with a set of bilinear equations satisfied by the $\tau$ functions in Gram determinant of the KP hierarchy,
and reduce them to bilinear equations satisfied by the 1D YO system (\ref{yoequation01})--(\ref{yoequation02}).
The reductions include mainly dimension reduction and complex conjugate reduction.
We should emphasize here that the most crucial and difficult issue is to find a general algebraic expression for the element of determinant such that the dimension reduction can be realized.
In Section 3, the dynamical behaviors of fundamental and higher-order rogue wave solutions are illustrated for different choices of free parameters.
The paper is concluded in Section 4 by a brief summary and discussion.

\section{Derivation of general rogue wave solutions}
This section is the core of the present paper, in which an explicit expression for general rogue wave solutions of the 1D YO system (\ref{yoequation01})--(\ref{yoequation02}) will be derived by Hirota's bilinear method.
To this end, let us first introduce dependent variable transformations
\begin{eqnarray}
\label{transformation01} S= e^{{\rm i} [\alpha x + (h+\alpha^2) t]  }\frac{g}{f},\ \ L= h- 2 \frac{\partial^2}{\partial x^2}\log f,
\end{eqnarray}
where $f$ is a real-valued function, $g$ is a complex-valued function and $\alpha$ and $h$ are real constants.
Then the 1D YO system (\ref{yoequation01})-(\ref{yoequation02}) is converted into the following bilinear
equations
\begin{eqnarray}
\label{yobilinearequation01} &&(D^2_x + 2{\rm i}\alpha D_x -{\rm i}D_t ) g \cdot f=0,\\
\label{yobilinearequation02} &&(D_xD_t+4)f \cdot f = 4gg^*,
\end{eqnarray}
where ${}^*$ denotes the complex
conjugation hereafter and
the $D$ is Hirota's bilinear differential operator defined by
\begin{eqnarray*}
D^n_xD^m_t(a\cdot b)=\bigg( \frac{\partial}{\partial x} - \frac{\partial}{\partial x'} \bigg)^n
\bigg( \frac{\partial}{\partial t} - \frac{\partial}{\partial t'} \bigg)^m
a(x,t)b(x',t')\bigg|_{x=x',t=t'} .
\end{eqnarray*}
Prior to the tedious process in deriving the polynomial solutions of the functions $f$ and $g$,
we highlight the main steps of the detailed derivation, as shown in the the subsequent subsections.

Firstly, we start from the following bilinear equations of the KP hierarchy:
\begin{eqnarray}
\label{kpeq01} && (D^2_{x_1} +2aD_{x_1} - D_{x_2})\tau_{n+1}\cdot \tau_n =0,\\
\label{kpeq02} && \left(\frac{1}{2}D_{x_1}D_{t_a}-1\right)\tau_n\cdot\tau_n = -\tau_{n+1}\tau_{n-1},
\end{eqnarray}
which admit a wide class of solutions in terms of Gram or Wronski determinant.
Among these determinant solutions, we need to look for algebraic solutions to satisfy the reduction condition:
\begin{eqnarray}
\label{reductioncondition01}(\partial_{x_2} + 2{\rm i}\partial_{t_a})\tau_n = c \tau_n,
\end{eqnarray}
such that these algebraic solutions satisfy the (1+1)-dimensional bilinear equations:
\begin{eqnarray}
\label{kpreductioneq01}&& (D^2_{x_1} +2aD_{x_1} - D_{x_2})\tau_{n+1}\cdot \tau_n =0,\\
\label{kpreductioneq02}&& (\rm i D_{x_1}D_{x_2}-4)\tau_n\cdot\tau_n = -4 \tau_{n+1}\tau_{n-1}.
\end{eqnarray}
Furthermore, by introducing the variable transformations:
\begin{eqnarray}
x_1=x,\ \ x_2=-{\rm i}t,
\end{eqnarray}
and taking $f=\tau_0$, $g=\tau_{1}$, $h=\tau_{-1}$, and $a={\rm i}\alpha$,
the above bilinear equations (\ref{kpreductioneq01})-(\ref{kpreductioneq02}) become
\begin{eqnarray}
\label{bilinereqgh01}&&(D^2_x + 2{\rm i}\alpha D_x - {\rm i}D_t) g \cdot f=0,\\
\label{bilinereqgh02}&&(D_xD_t+4)f \cdot f = 4gh.
\end{eqnarray}
Lastly, by requiring the real and complex conjugation condition:
\begin{eqnarray}
f=\tau_0: \mbox{real}, \ \ g=\tau_{1},\ \ h=\tau_{-1}=g^*,
\end{eqnarray}
in the algebraic solutions, then the bilinear equations (\ref{bilinereqgh01})-(\ref{bilinereqgh02}) are reduced to the bilinear equations (\ref{yobilinearequation01})--(\ref{yobilinearequation02}), hence the general higher-order rogue wave solutions are obtained through
the reductions.

\subsection{Gram determinant solution for the bilinear equations in KP hierarchy}

In this subsection, through the Lemma below, we present and prove the a pair of bilinear equations satisfied by the $\tau$ functions of the KP hierarchy.

\textbf{Lemma 2.1}
Let $m^{(n)}_{ij}$, depending on $\varphi^{(n)}_i$ and $\psi^{(n)}_j$, be function of the variables $x_1$, $x_2$ and $t_a$,
and satisfy the following differential and difference relations:
\begin{eqnarray}\label{diff-rule-01}
\nonumber && \partial_{x_1} m^{(n)}_{ij} = \varphi^{(n)}_i \psi^{(n)}_j,\\
\nonumber && \partial_{x_2} m^{(n)}_{ij} = [\partial_{x_1}\varphi^{(n)}_i] \psi^{(n)}_j - \varphi^{(n)}_i [\partial_{x_1}\psi^{(n)}_j],\\
&& \partial_{t_a} m^{(n)}_{ij} = -\varphi^{(n-1)}_i \psi^{(n+1)}_j,\\
\nonumber &&  m^{(n+1)}_{ij} =  m^{(n)}_{ij} + \varphi^{(n)}_i \psi^{(n+1)}_j,
\end{eqnarray}
where $\varphi^{(n)}_i$ and $\psi^{(n)}_j$ are functions satisfying
\begin{eqnarray}\label{diff-rule-02}
\partial_{x_2}\varphi^{(n)}_i = \partial^2_{x_1}\varphi^{(n)}_i,\ \
\varphi^{(n+1)}_i = (\partial_{x_1}-a)\varphi^{(n)}_i,\ \
\partial_{x_2}\psi^{(n)}_j = -\partial^2_{x_1}\psi^{(n)}_j,\ \
\psi^{(n-1)}_i = -(\partial_{x_1}+a)\psi^{(n)}_j.
\end{eqnarray}
Then the $\tau$ functions of the following determinant form
\begin{eqnarray}
\label{taufunction01}\tau_n=\det_{1\leq i,j \leq N} (m^{(n)}_{ij}),
\end{eqnarray}
satisfy the following bilinear equations (\ref{kpeq01}) and (\ref{kpeq02}) in KP hierarchy:
\begin{eqnarray}
\label{b-kp-01} && (D^2_{x_1} + 2a D_{x_1} -D_{x_2})\tau_{n+1}\cdot \tau_n =0,\\
\label{b-kp-02} && (\frac{1}{2}D_{x_1}D_{t_a}-1)\tau_n\cdot\tau_n = -\tau_{n+1}\tau_{n-1}.
\end{eqnarray}

\emph{Proof:}
By using the differential formula of determinant
\begin{eqnarray}\label{ddet-01}
\partial_x \det_{1\leq i,j \leq N} (a_{ij}) = \sum^N_{i,j=1}\Delta_{ij} \partial_x a_{ij},
\end{eqnarray}
and the expansion formula of bordered determinant
\begin{eqnarray}
\det\left( \begin{array}{cc} a_{ij} & b_i \\ c_j & d \end{array} \right)
=-\sum^N_{i,j}\Delta_{ij} b_i c_j + d\det(a_{ij}),
\end{eqnarray}
with $\Delta_{ij}$ being the $(i,j)$-cofactor of the matrix $(a_{ij})$,
one can check that
the derivatives and shifts of the $\tau$ function are expressed by
the bordered determinants as follows:
\begin{eqnarray*}
&&
\partial_{x_1}\tau_n=\begin{vmatrix} m^{(n)}_{ij} & \varphi^{(n)}_i \\ -\psi^{(n)}_j & 0 \end{vmatrix},\ \
\partial^2_{x_1}\tau_n=\begin{vmatrix} m^{(n)}_{ij} & \partial_{x_1}\varphi^{(n)}_i \\ -\psi^{(n)}_j & 0 \end{vmatrix}
+ \begin{vmatrix} m^{(n)}_{ij} & \varphi^{(n)}_i \\ -\partial_{x_1}\psi^{(n)}_j & 0 \end{vmatrix},
\\
&&
\partial_{x_2}\tau_n=\begin{vmatrix} m^{(n)}_{ij} & \partial_{x_1}\varphi^{(n)}_i \\ -\psi^{(n)}_j & 0 \end{vmatrix}
- \begin{vmatrix} m^{(n)}_{ij} & \varphi^{(n)}_i \\ -\partial_{x_1}\psi^{(n)}_j & 0 \end{vmatrix},\ \
\partial_{t_a}\tau_n=\begin{vmatrix} m^{(n)}_{ij} & \varphi^{(n-1)}_i \\ \psi^{(n+1)}_j & 0 \end{vmatrix},
\\
&&
(\partial_{x_1}\partial_{t_a}-1)\tau_n=\begin{vmatrix} m^{(n)}_{ij} & \varphi^{(n-1)}_i & \varphi^{(n)}_i \\ \psi^{(n+1)}_j & 0 & -1 \\ -\psi^{(n)}_j & -1 & 0 \end{vmatrix},
\\
&&
\tau_{n+1}=\begin{vmatrix} m^{(n)}_{ij} & \varphi^{(n)}_i \\ -\psi^{(n+1)}_j & 1 \end{vmatrix},\ \
\tau_{n-1}=\begin{vmatrix} m^{(n)}_{ij} & \varphi^{(n-1)}_i \\ \psi^{(n)}_j & 1 \end{vmatrix}, \ \
(\partial_{x_1}+a)\tau_{n+1}=\begin{vmatrix} m^{(n)}_{ij} & \partial_{x_1}\varphi^{(n)}_i \\ -\psi^{(n+1)}_j & a \end{vmatrix},
\\
&&
(\partial_{x_1}+a)^2\tau_{n+1}= \begin{vmatrix} m^{(n)}_{ij} & \partial^2_{x_1}\varphi^{(n)}_i \\ -\psi^{(n+1)}_j & a^2 \end{vmatrix}
+\begin{vmatrix} m^{(n)}_{ij} & \partial_{x_1}\varphi^{(n)}_i & \varphi^{(n)}_i \\ -\psi^{(n+1)}_j & a & 1 \\ -\psi^{(n)}_j & 0 & 0 \end{vmatrix},
\\
&&
(\partial_{x_2}+a^2)\tau_{n+1}= \begin{vmatrix} m^{(n)}_{ij} & \partial^2_{x_1}\varphi^{(n)}_i \\ -\psi^{(n+1)}_j & a^2 \end{vmatrix}
-\begin{vmatrix} m^{(n)}_{ij} & \partial_{x_1}\varphi^{(n)}_i & \varphi^{(n)}_i \\ -\psi^{(n+1)}_j & a & 1 \\ -\psi^{(n)}_j & 0 & 0 \end{vmatrix}.
\end{eqnarray*}
With the help of these relations, one has
\begin{eqnarray}
\label{rw-1d-yo-J1}\nonumber && \hspace{0.3cm} (\partial_{x_1}\partial_{t_a}-1)\tau_n \times  \tau_n - \partial_{x_1}\tau_{n} \times \partial_{t_a}\tau_{n}
+(-\tau_{n-1})(-\tau_{n+1}) \\
&& = \begin{vmatrix} m^{(n)}_{ij} & \varphi^{(n-1)}_i & \varphi^{(n)}_i \\ \psi^{(n+1)}_j & 0 & -1 \\ -\psi^{(n)}_j & -1 & 0 \end{vmatrix}
\times \begin{vmatrix} m^{(n)}_{ij}  \end{vmatrix}
-
\begin{vmatrix} m^{(n)}_{ij} & \varphi^{(n)}_i \\ -\psi^{(n)}_j & 0 \end{vmatrix}
\times \begin{vmatrix} m^{(n)}_{ij} & \varphi^{(n-1)}_i \\ \psi^{(n+1)}_j & 0 \end{vmatrix}
+
\begin{vmatrix} m^{(n)}_{ij} & \varphi^{(n-1)}_i \\ -\psi^{(n)}_j & -1 \end{vmatrix}
\times \begin{vmatrix} m^{(n)}_{ij} & \varphi^{(n)}_i \\ \psi^{(n+1)}_j & -1 \end{vmatrix},\\
\label{rw-1d-yo-J2} \nonumber && \hspace{0.3cm} \frac{1}{2}(\partial^2_{x_1}+2a\partial_{x_1} -\partial_{x_2}) \tau_{n+1}\times \tau_n - (\partial_{x_1}+a)\tau_{n+1} \times \partial_{x_1}\tau_{n} + \tau_{n+1}\times \frac{1}{2}(\partial^2_{x_1} + \partial_{x_2})\tau_n \\
&& = \begin{vmatrix} m^{(n)}_{ij} & \partial_{x_1}\varphi^{(n)}_i & \varphi^{(n)}_i \\ -\psi^{(n+1)}_j & a & 1 \\ -\psi^{(n)}_j & 0 & 0 \end{vmatrix}
\times \begin{vmatrix} m^{(n)}_{ij}  \end{vmatrix}
-\begin{vmatrix} m^{(n)}_{ij} & \varphi^{(n)}_i \\ -\psi^{(n)}_j & 0 \end{vmatrix}
\times \begin{vmatrix} m^{(n)}_{ij} & \partial_{x_1}\varphi^{(n)}_i \\ -\psi^{(n+1)}_j & a \end{vmatrix}
+\begin{vmatrix} m^{(n)}_{ij} & \partial_{x_1}\varphi^{(n)}_i \\ -\psi^{(n)}_j & 0 \end{vmatrix}
\times \begin{vmatrix} m^{(n)}_{ij} & \varphi^{(n)}_i \\ -\psi^{(n+1)}_j & 1 \end{vmatrix}.\ \ \ \ \ \
\end{eqnarray}
The r.h.s of both (\ref{rw-1d-yo-J1}) and (\ref{rw-1d-yo-J2}) are identically zero because of the
Jacobi identity and hence
the $\tau$ functions (\ref{taufunction01}) satisfy the bilinear equations (\ref{b-kp-01}) and (\ref{b-kp-02}).
This completes the proof.

\subsection{Algebraic solutions for the (1+1)-dimensional YO system}
This subsection is crucial in the KP-hierarchy reductions.
We will construct an algebraic expression for the elements of $\tau$ function of preceding subsection so that the dimension reduction condition (\ref{reductioncondition01}) is satisfied.
The main result is given by the following Lemma.

\textbf{Lemma 2.2} Suppose the entries $m_{kl}^{(\nu\mu n)}$ of the matrix $m$ are
\begin{eqnarray}
m_{kl}^{(\nu\mu n)}
=\left.(A_k^{(\nu)}B_l^{(\mu)}m^{(n)})\right|_{p=\zeta,q=\zeta^*},
\end{eqnarray}
where
\begin{eqnarray*}
&& m^{(n)}=\frac{1}{p+q}\left(-\frac{p-a}{q+a}\right)^ne^{\xi+\eta},\ \
 \xi=px_1+p^2x_2,\ \
\eta=qx_1-q^2x_2,\\
&& \zeta=\zeta_r + {\rm i}\zeta_i,\ \
 \zeta_r= \pm\frac{\sqrt{3}}{12}\frac{K^2-4\alpha^2}{K},\ \ \zeta_i=\frac{1}{12}K + \frac{1}{3}\frac{\alpha^2}{K} + \frac{2}{3}\alpha,
\end{eqnarray*}
\begin{equation*}
K= \left( 8\alpha^3 +108 + 12 \sqrt{12\alpha^3+81} \right)^{1/3}, \ \ \  (\alpha > -\frac{3}{2}2^{1/3}),
\end{equation*}
and $A_k^{(\nu)}$ and $B_l^{(\mu)}$ are differential operators with respect to $p$ and $q$, respectively, given by
\begin{eqnarray}
&& A_n^{(\nu)}=
\sum_{k=0}^na_k^{(\nu)}\frac{[(p-a)\partial_p]^{n-k}}{(n-k)!},\quad
 n \geq 0,
\\
&& B_n^{(\nu)}=
\sum_{k=0}^nb_k^{(\nu)}\frac{[(q+a)\partial_q]^{n-k}}{(n-k)!},\quad
 n\geq0,
\end{eqnarray}
where $a_k^{(\nu)}$ and $b_k^{(\nu)}$ are constants satisfying the iterated relations
\begin{eqnarray}
&& a^{(\nu+1)}_{k}=\sum^{k}_{j=0} \frac{2^{j+2} (p-a)^2 + (-1)^{j}  \frac{2{\rm i}}{p-a} +2a(p-a)}{ (j+2)!} a^{(\nu)}_{k-j},
\ \ \nu=0,1,2,\cdots,\\
&& b^{(\nu+1)}_{k}=\sum^{k}_{j=0} \frac{2^{j+2} (q+a)^2 - (-1)^{j}  \frac{2{\rm i}}{q+a} -2a(q+a)}{ (j+2)!} b^{(\nu)}_{k-j},
\ \ \nu=0,1,2,\cdots\,,
\end{eqnarray}
then the determinant
\begin{eqnarray}
\label{tau_algebra}
\tau_n=\det_{1\leq i,j\leq N}\left(m^{(N-i,N-j,n)}_{2i-1,2j-1}\right)
=\begin{vmatrix}
m_{11}^{(N-1,N-1,n)} & m_{13}^{(N-1,N-2,n)} &\cdots & m_{1,2N-1}^{(N-1,0,n)} \\
m_{31}^{(N-2,N-1,n)} & m_{33}^{(N-2,N-2,n)} &\cdots & m_{3,2N-1}^{(N-2,0,n)} \\
\vdots &\vdots & & \vdots \\
m_{2N-1,1}^{(0,N-1,n)} &m_{2N-1,3}^{(0,N-2,n)} & \cdots & m_{2N-1,2N-1}^{(0,0,n)}
\end{vmatrix},
\end{eqnarray}
satisfies the bilinear equations
\begin{eqnarray}
\label{1d-eq-01} && (D^2_{x_1} +2aD_{x_1} - D_{x_2})\tau_{n+1}\cdot \tau_n =0,\\
\label{1d-eq-02} && ({\rm i}D_{x_1}D_{x_2}-4)\tau_n\cdot\tau_n = -4\tau_{n+1}\tau_{n-1}.
\end{eqnarray}

\emph{Proof.}
Firstly, we introduce the functions $\tilde{m}^{(n)}$, $\tilde{\varphi}^{(n)}$ and $\tilde{\psi}^{(n)}$ of the form
\begin{eqnarray*}
\tilde{m}^{(n)} = \frac{1}{p+q} \left(-\frac{p-a}{q+a}\right)^n e^{\tilde{\xi}+\tilde{\eta}},\ \
\tilde{\varphi}^{(n)}=(p-a)^ne^{\tilde{\xi}},\ \ \tilde{\psi}^{(n)}=\left(-\frac{1}{q+a}\right)^{n}e^{\tilde{\eta}},
\end{eqnarray*}
where
\begin{eqnarray*}
\tilde{\xi}=\frac{1}{p-a}t_a +px_1 + p^2 x_2,\ \
\tilde{\eta}=\frac{1}{q+a}t_a +qx_1 - q^2 x_2.
\end{eqnarray*}
These functions satisfy the differential and difference rules:
\begin{eqnarray*}
&& \partial_{x_1} \tilde{m}^{(n)} = \tilde{\varphi}^{(n)} \tilde{\psi}^{(n)},\\
&& \partial_{x_2} \tilde{m}^{(n)} = [\partial_{x_1}\tilde{\varphi}^{(n)}] \tilde{\psi}^{(n)} - \tilde{\varphi}^{(n)} [\partial_{x_1}\tilde{\psi}^{(n)}],\\
&& \partial_{t_a} \tilde{m}^{(n)} = -\tilde{\varphi}^{(n-1)} \tilde{\psi}^{(n+1)},\\
&&  \tilde{m}^{(n+1)} =  \tilde{m}^{(n)} + \tilde{\varphi}^{(n)} \tilde{\psi}^{(n+1)},
\end{eqnarray*}
and
\begin{eqnarray*}
\partial_{x_2}\tilde{\varphi}^{(n)} = \partial^2_{x_1}\tilde{\varphi}^{(n)},\ \
\tilde{\varphi}^{(n+1)} = (\partial_{x_1}-a)\tilde{\varphi}^{(n)},\ \
\partial_{x_2}\tilde{\psi}^{(n)} = -\partial^2_{x_1}\tilde{\psi}^{(n)},\ \
\tilde{\psi}^{(n-1)} = -(\partial_{x_1}+a)\tilde{\psi}^{(n)}.
\end{eqnarray*}
We then define
\begin{eqnarray}
\tilde{m}^{(\nu\mu n)}_{ij} = A_i^{(\nu)}B_j^{(\mu)}\tilde{m}^{(n)},\ \
\tilde{\varphi}^{(\nu n)}_i = A_i^{(\nu)} \tilde{\varphi}^{(n)},\ \
\tilde{\psi}^{(\mu n)}_j = B_j^{(\mu)} \tilde{\psi}^{(n)}\,.
\end{eqnarray}
Since the operators $A_i^{(\nu)}$ and $B_j^{(\mu)}$ commute with differential operators $\partial_{x_1}$, $\partial_{x_2}$
and $\partial_{t_a}$, these functions $\tilde{m}^{(\nu\mu n)}_{ij}$, $\tilde{\varphi}^{(\nu n)}_i$ and $\tilde{\psi}^{(\mu n)}_j$ obey the differential and difference relations as well
(\ref{diff-rule-01})-(\ref{diff-rule-02}).
From Lemma 2.1, we know that for an arbitrary sequence of indices $(i_1, i_2, \ldots, i_N ; \nu_1, \nu_2, \ldots, \nu_N ; j_1, j_2,\ldots , j_N ; \mu_1, \mu_2, \ldots, \mu_N)$,
the determinant
\begin{eqnarray*}
\tilde{\tau}_n = \det_{1\leq k,l\leq N}(\tilde{m}^{(\nu_k,\mu_l,n)}_{i_k,j_l})
\end{eqnarray*}
satisfies the bilinear equations (\ref{b-kp-01}) and (\ref{b-kp-02}),
for instance, the bilinear equations (\ref{b-kp-01}) and (\ref{b-kp-02}) hold for $\tilde{\tau}_n = \det_{1\leq i,j\leq N}(\tilde{m}^{(N-i,N-j,n)}_{2i-1,2j-1})$ with arbitrary parameters $p$ and $q$.
Based on the Leibniz rule, one has,
\begin{eqnarray}
\nonumber  [(p-a)\partial_p]^m\left(p^2+ \frac{2{\rm i}}{p-a}\right) &=& \sum^m_{l=0}\left( \begin{array}{c} m \\ l \end{array} \right) \left[2^l (p-a)^2 + (-1)^l\frac{2{\rm i}}{p-a} +2a(p-a) \right][(p-a)\partial_p]^{m-l} \\
 && + a^2[(p-a)\partial_p]^{m},
\end{eqnarray}
and
\begin{eqnarray}
\nonumber  [(q+a)\partial_q]^m\left(q^2- \frac{2{\rm i}}{q+a}\right) &=& \sum^m_{l=0}\left( \begin{array}{c} m \\ l \end{array} \right) \left[2^l (q+a)^2 - (-1)^l\frac{2{\rm i}}{q+a} - 2a(q+a) \right][(q+a)\partial_q]^{m-l} \\
 && + a^2[(q+a)\partial_q]^{m}\,.
\end{eqnarray}
Furthermore, one can derive
\begin{eqnarray*}
&& \left[A^{(\nu)}_n,p^2+\frac{2{\rm i}}{p-a}\right]
=\sum_{k=0}^{n-1}\frac{a_k^{(\nu)}}{(n-k)!}
\left[((p-a)\partial_p)^{n-k},p^2+\frac{2{\rm i}}{p-a}\right] \\
&& =\sum_{k=0}^{n-1}\frac{a_k^{(\nu)}}{(n-k)!}
\sum^{n-k}_{l=1}\left( \begin{array}{c} n-k \\ l \end{array} \right)
\left(2^l (p-a)^2 + (-1)^l\frac{2{\rm i}}{p-a} +2a(p-a) \right)
((p-a)\partial_p)^{n-k-l},
\end{eqnarray*}
where $[\ ,\ ]$ is the commutator defined by $[X,Y]=XY-YX$.

Let $\zeta$ be the solution of the cubic equation
$$
2(p-a)^2-\frac{2{\rm i}}{p-a}+2a(p-a)=0,
$$
with $a={\rm i}\alpha$, then $\zeta$ has an explicit expression given previously.
Hence we have
$$
\left.\left[A^{(\nu)}_n,p^2+\frac{2{\rm i}}{p-a}\right]\right|_{p=\zeta}=0,
$$
for $n=0,1$ and
\begin{eqnarray*}
&&\left.\left[A^{(\nu)}_n,p^2+\frac{2{\rm i}}{p-a}\right]\right|_{p=\zeta} \\
&&=\left.\sum_{k=0}^{n-2}\frac{a_k^{(\nu)}}{(n-k)!}
\sum^{n-k}_{l=2}\left( \begin{array}{c} n-k \\ l \end{array} \right)
\left\{2^l (p-a)^2 + (-1)^l\frac{2{\rm i}}{p-a} +2a(p-a) \right\}
((p-a)\partial_p)^{n-k-l}\right|_{p=\zeta} \\
&&=\left.\sum_{k=0}^{n-2}\sum^{n-k-2}_{j=0}
\frac{a^{(\nu)}_k}{(j+2)!(n-k-j-2)!}
\left\{2^{j+2} (p-a)^2 + (-1)^{j} \frac{2{\rm i}}{p-a} +2a(p-a) \right\}
((p-a)\partial_p)^{n-k-j-2}\right|_{p=\zeta} \\
&&=\left.\sum^{n-2}_{\hat{k}=0} \left( \sum^{\hat{k}}_{\hat{j}=0} \frac{2^{\hat{j}+2} (p-a)^2 + (-1)^{\hat{j}}  \frac{2{\rm i}}{p-a} +2a(p-a)}{ (\hat{j}+2)!} a^{(\nu)}_{\hat{k}-\hat{j}} \right) \frac{((p-a)\partial_p)^{n-2-\hat{k}} }{(n-2-\hat{k})!}\right|_{p=\zeta} \\
&&=\left.\sum^{n-2}_{\hat{k}=0} a^{(\nu+1)}_{\hat{k}} \frac{((p-a)\partial_p)^{n-2-\hat{k}} }{(n-2-\hat{k})!}\right|_{p=\zeta} \\
&&=\left.  A^{(\nu+1)}_{n-2}\right|_{p=\zeta},
\end{eqnarray*}
for $n\ge2$.
Thus the differential operator $A_n^{(\nu)}$ satisfies the following relation
\begin{eqnarray}
&&\left. \left[A_n^{(\nu)},p^2+\frac{2{\rm i}}{p-a}\right]\right|_{p=\zeta}
=\left.A_{n-2}^{(\nu+1)}\right|_{p=\zeta},
\end{eqnarray}
where we define $A_n^{(\nu)}=0$ for $n<0$.

Similarly, it is shown that the differential operator $B_n^{(\nu)}$ satisfies
\begin{eqnarray}
&&\left. \left[B_n^{(\nu)},q^2-\frac{2{\rm i}}{q+a}\right]\right|_{q=\zeta^*}
=\left.B_{n-2}^{(\nu+1)}\right|_{q=\zeta^*},
\end{eqnarray}
where we define $B_n^{(\nu)}=0$ for $n<0$. \\

Consequently, by referring to above two relations,  we have
\begin{eqnarray*}
&& \left. (\partial_{x_2}+2{\rm i}\partial_{t_a})\tilde{m}_{kl}^{(\nu\mu n)}\right|_{p=\zeta,q=\zeta^*}
\\
&=& \left. \left[A_k^{(\nu)}B_l^{(\mu)}
(\partial_{x_2}+2{\rm i}\partial_{t_a})\tilde{m}^{(n)}\right]\right|_{p=\zeta,q=\zeta^*}
\\
&=& \left.\left(A_k^{(\nu)}B_l^{(\mu)}
\left(p^2-q^2+2{\rm i} \left(\frac{1}{p-a}+\frac{1}{q+a}\right)\right) \tilde{m}^{(n)}\right)\right|_{p=\zeta,q=\zeta^*}
\\
&=& \left.\left(A_k^{(\nu)}\left(p^2+\frac{2{\rm i} }{p-a}\right)B_l^{(\mu)}
\tilde{m}^{(n)}\right)\right|_{p=\zeta,q=\zeta^*}
-\left.\left(A_k^{(\nu)}B_l^{(\mu)}\left(q^2-\frac{2{\rm i} }{q+a}\right)
\tilde{m}^{(n)}\right)\right|_{p=\zeta,q=\zeta^*}
\\
&=& \left.\left\{\left(\left(p^2+\frac{2{\rm i}}{p-a}\right)A_k^{(\nu)}+A_{k-2}^{(\nu+1)}\right)B_l^{(\mu)}
\tilde{m}^{(n)}\right\}\right|_{p=\zeta,q=\zeta^*}
\\
&&
-\left.\left\{A_k^{(\nu)}\left(\left(q^2-\frac{2{\rm i}}{q+a}\right)B_l^{(\mu)} + B_{l-2}^{(\mu+1)}\right)
\tilde{m}^{(n)}\right\}\right|_{p=\zeta,q=\zeta^*}
\\
&=& \left(\zeta^2+\frac{2{\rm i}}{\zeta-{\rm i}\alpha}\right) \left.\tilde{m}_{kl}^{(\nu\mu n)}\right|_{p=\zeta,q=\zeta^*}
+\left.\tilde{m}_{k-2,l}^{(\nu+1,\mu, n)}\right|_{p=\zeta,q=\zeta^*}
-\left(\zeta^{*2}-\frac{2{\rm i}}{\zeta^*+{\rm i}\alpha}\right) \left.\tilde{m}_{kl}^{(\nu\mu n)}\right|_{p=\zeta,q=\zeta^*}
-\left.\tilde{m}_{k,l-2}^{(\nu,\mu+1,n)}\right|_{p=\zeta,q=\zeta^*} .
\end{eqnarray*}
By using the formula (\ref{ddet-01}) and the above relation, the differential of the following determinant
\begin{eqnarray*}
\tilde{\tilde{\tau}}_n=\det_{1\leq i,j\leq N}\left(\left.\tilde{m}_{2i-1,2j-1}^{(N-i,N-j, n)}\right|_{p=\zeta,q=\zeta^*}\right)
\end{eqnarray*}
can be calculated as
\begin{eqnarray*}
&&(\partial_{x_2}+2{\rm i} \partial_{t_a})\tilde{\tilde{\tau}}_n \\
&=& \sum^N_{i=1} \sum^N_{j=1} \Delta_{ij} (\partial_{x_2}+2{\rm i} \partial_{t_a})\left( \left. \tilde{m}^{(N-i,N-j,n)}_{2i-1,2j-1}\right|_{p=\zeta,q=\zeta^*} \right)\\
&=& \sum^N_{i=1} \sum^N_{j=1} \Delta_{ij}
\left[\left(\zeta^2+\frac{2{\rm i}}{\zeta-{\rm i}\alpha}\right)  \left. \tilde{m}^{(N-i,N-j,n)}_{2i-1,2j-1}\right|_{p=\zeta,q=\zeta^*} + \left. m^{(N-i+1,N-j,n)}_{2i-3,2j-1}\right|_{p=\zeta,q=\zeta^*}  \right.
\\
&& \left.
- \left(\zeta^{*2}-\frac{2{\rm i}}{\zeta^*+{\rm i}\alpha}\right)  \left. \tilde{m}^{(N-i,N-j,n)}_{2i-1,2j-1}\right|_{p=\zeta,q=\zeta^*}
- \left. m^{(N-i,N-j+1,n)}_{2i-1,2j-3}\right|_{p=\zeta,q=\zeta^*}  \right]\\
&=& \left(\zeta^2+\frac{2{\rm i}}{\zeta-{\rm i}\alpha}\right) N \tilde{\tilde{\tau}}_n
+  \sum^N_{i=1} \sum^N_{j=1} \Delta_{ij} \left. \tilde{m}^{(N-i+1,N-j,n)}_{2i-3,2j-1}\right|_{p=\zeta,q=\zeta^*}
 - \left(\zeta^{*2}-\frac{2{\rm i}}{\zeta^*+{\rm i}\alpha}\right) N \tilde{\tilde{\tau}}_n
 -  \sum^N_{i=1} \sum^N_{j=1} \Delta_{ij} \left. \tilde{m}^{(N-i,N-j+1,n)}_{2i-1,2j-3}\right|_{p=\zeta,q=\zeta^*},
\end{eqnarray*}
where $\Delta_{ij}$ is the $(i,j)$-cofactor of the matrix $\left(\left.\tilde{m}_{2i-1,2j-1}^{(N-i,N-j, n)}\right|_{p=\zeta,q=\zeta^*} \right)_{1\leq i,j\leq N}$.
For the term $\sum^N_{i=1} \sum^N_{j=1} \Delta_{ij} \left. \tilde{m}^{(N-i+1,N-j,n)}_{2i-3,2j-1}\right|_{p=\zeta,q=\zeta^*}$,
it vanishes, since for $i=1$ this summation is a determinant with the elements in first row being zero
and for $i=2,3,\ldots$ this summation is a determinant with two identical rows.
Similarly, the term $\sum^N_{i=1} \sum^N_{j=1} \Delta_{ij} \left. \tilde{m}^{(N-i,N-j+1,n)}_{2i-1,2j-3}\right|_{p=\zeta,q=\zeta^*}$ vanishes.
Therefore, $\tilde{\tilde{\tau}}_n$ satisfies the reduction condition
\begin{eqnarray}\label{red-condition}
(\partial_{x_2}+2{\rm i} \partial_{t_a})\tilde{\tilde{\tau}}_n = \left(\zeta^2-\zeta^{*2}+\frac{2{\rm i}}{\zeta-{\rm i}\alpha}+\frac{2{\rm i}}{\zeta^*+{\rm i}\alpha}\right) N \tilde{\tilde{\tau}}_n.
\end{eqnarray}
Since $\tilde{\tilde{\tau}}_n$ is a special case of $\tilde{\tau}_n$,
it also satisfies the bilinear equations (\ref{b-kp-01}) and (\ref{b-kp-02}) with $\tau_n$ replaced by $\tilde{\tilde{\tau}}_n$.
From (\ref{b-kp-01}), (\ref{b-kp-02}) and (\ref{red-condition}), it is obvious that $\tilde{\tilde{\tau}}_n$ satisfies the (1+1)-dimensional bilinear equations
\begin{eqnarray}
&& (D^2_{x_1} +2aD_{x_1} - D_{x_2})\tilde{\tilde{\tau}}_{n+1}\cdot \tilde{\tilde{\tau}}_n =0,\\
&& ({\rm i} D_{x_1}D_{x_2}-4)\tilde{\tilde{\tau}}_n\cdot\tilde{\tilde{\tau}}_n = -4\tilde{\tilde{\tau}}_{n+1}\tilde{\tilde{\tau}}_{n-1}.
\end{eqnarray}
Due to the reduction condition (\ref{red-condition}), $t_a$ becomes a dummy variable which can be taken as zero. Thus $\left.\tilde{m}_{2i-1,2j-1}^{(N-i,N-j, n)}\right|_{p=\zeta,q=\zeta^*}$ and $\tilde{\tilde{\tau}}_n$
reduce to $m_{2i-1,2j-1}^{(N-i,N-j,n)}$ and $\tau_n$ (\ref{tau_algebra}) in Lemma 2.2. Therefore $\tau_n$ satisfies the bilinear equations (\ref{1d-eq-01}) and (\ref{1d-eq-02}) and the proof is complete.
\subsection{Complex conjugate condition and regularity }
From Lemma 2.2, by taking the independent variable transformations:
\begin{eqnarray}
x_1=x,\ \ x_2=-{\rm i}t,\ \
\end{eqnarray}
it is found that $f=\tau_0$, $g=\tau_{1}$ and $h=\tau_{-1}$
satisfy the (1+1)-dimensional bilinear equations:
\begin{eqnarray}
&&(D^2_x + 2{\rm i}\alpha D_x -{\rm i}D_t ) g \cdot f=0,\\
&&(D_xD_t+4)f \cdot f = 4gh\,.
\end{eqnarray}
Next we consider the complex conjugate condition and the regularity
(non-singularity) of solutions.
The complex conjugate condition requires
\begin{eqnarray}
\tau_0: \mbox{real}, \ \ \tau_{-1}=\tau^*_1.
\end{eqnarray}
Since $x_1 =x$ is real and $x_2=-{\rm i}t$ is pure imaginary, the complex conjugate condition can be easily satisfied by taking the parameters $a^{(0)}_k$ and $b^{(0)}_k$ to be complex
conjugate to each other.
It then follows
\begin{eqnarray}\label{complex-conj-1}
\left.b^{(\nu)}_k\right|_{q=\zeta^*}
=\left(\left.a^{(\nu)}_k\right|_{p=\zeta}\right)^*,
\end{eqnarray}
for $\nu=0,1,2,\ldots$.
Then, by referring to (\ref{complex-conj-1}), we have
\begin{eqnarray}
m^{(\nu, \mu, n)*}_{ij}
=\left.m^{(\nu, \mu, n)}_{ij}\right|_{a^{(\nu)}_k\leftrightarrow b^{(\nu)}_k,x_2\leftrightarrow-x_2,a\leftrightarrow -a,\zeta\leftrightarrow \zeta^*}= m^{(\mu, \nu, -n)}_{ji},
\end{eqnarray}
which implies
\begin{eqnarray}
\tau^*_n  = \tau_{-n}.
\end{eqnarray}
On the other hand, under the condition (\ref{complex-conj-1}), we can show that $\tau_0$ is non-zero for all $(x,t)$.
Note that $f=\tau_0$ is the determinant of a Hermitian matrix $M=\left(m^{(N-i,N-j,0)}_{2i-1,2j-1}\right)_{1\leq i,j\leq N}$.
From the appendix in \cite{ohta2013dynamics}, it is known that
when the real part of the parameter $\zeta$ is positive, the element of the Hermitian matrix $M$ can be
written as an integral
\begin{eqnarray}
m^{(N-i,N-j,0)}_{2i-1,2j-1}= \left. \int^x_{-\infty} A_{2i-1}^{(N-i)}B_{2j-1}^{(N-j)} e^{\xi+\eta}dx\right|_{p=\zeta,q=\zeta^*}.
\end{eqnarray}
For any non-zero column vector $\upsilon=(v_1,v_2,\ldots,v_N)^T$ and $\upsilon^{\dagger}$ being its complex transpose,
one can obtain
\begin{eqnarray*}
\upsilon^{\dagger} M \upsilon =\sum^N_{i,j=1} \upsilon^{\dagger}_i m^{(N-i,N-j,0)}_{2i-1,2j-1} \upsilon_j
= \int^x_{-\infty} \left|\sum^N_{i=1}\upsilon^{\dagger}_i   \left.A_{2i-1}^{(N-i)}  e^{\xi}\right|_{p=\zeta} \right|^2 dx >0,
\end{eqnarray*}
which shows that the Hermitian matrix $M$ is positive definite, hence the
denominator $f =\det M >0$.

When the real part of the parameter $p$ is negative, the element of the Hermitian matrix $M$ can be cast into
\begin{eqnarray}
m^{(N-i,N-j,0)}_{2i-1,2j-1}= - \left. \int^{+\infty}_x A_{2i-1}^{(N-i)}B_{2j-1}^{(N-j)} e^{\xi+\eta}dx\right|_{p=\zeta,q=\zeta^*}.
\end{eqnarray}
Then one obtains
\begin{eqnarray*}
\upsilon^{\dagger} M \upsilon =\sum^N_{i,j=1} \upsilon^{\dagger}_i m^{(N-i,N-j,0)}_{2i-1,2j-1} \upsilon_j
= - \int^{+\infty}_x \left|\sum^N_{i=1}\upsilon^{\dagger}_i   \left.A_{2i-1}^{(N-i)}  e^{\xi}\right|_{p=\zeta} \right|^2 dx <0,
\end{eqnarray*}
which proves that the Hermitian matrix $M$ is negative definite, hence the
denominator $f =\det M <0$.
Therefore, for either positive or negative of the parameter $\zeta$,  the rogue wave solution of the short wave and long wave components is always nonsingular.

To summarize the results, we have  the
following theorem for the general higher order rogue wave
solutions of the 1D YO system (\ref{yoequation01})-(\ref{yoequation02}):

\textbf{Theorem 2.3}
The 1D YO system (\ref{yoequation01})-(\ref{yoequation02}) has the non-singular rational solutions
\begin{eqnarray}\label{theoremeq01}
S= e^{{\rm i} [\alpha x + (h+\alpha^2) t]  }\frac{\tau_1}{\tau_0},\ \ L= h- 2 \frac{\partial^2}{\partial x^2}\log \tau_0,
\end{eqnarray}
with
\begin{eqnarray}\label{theoremeq02}
\tau_n=\det_{1\leq i,j\leq N}\left(m^{(N-i,N-j,n)}_{2i-1,2j-1}\right)
=\begin{vmatrix}
m_{11}^{(N-1,N-1,n)} & m_{13}^{(N-1,N-2,n)} &\cdots & m_{1,2N-1}^{(N-1,0,n)} \\
m_{31}^{(N-2,N-1,n)} & m_{33}^{(N-2,N-2,n)} &\cdots & m_{3,2N-1}^{(N-2,0,n)} \\
\vdots &\vdots & & \vdots \\
m_{2N-1,1}^{(0,N-1,n)} &m_{2N-1,3}^{(0,N-2,n)} & \cdots & m_{2N-1,2N-1}^{(0,0,n)}
\end{vmatrix},
\end{eqnarray}
where $g=\tau_1$, $f=\tau_0$ and $h=\tau_{-1}$, and the elements in determinant $\tau_n$  are defined by
\begin{eqnarray}\label{theoremeq03}
m^{(\nu,\mu,n)}_{i,j}= \left. \sum_{k=0}^i \sum_{l=0}^j  \frac{ a_k^{(\nu)} }{(i-k)!} \frac{ a_l^{(\mu)*} }{(j-l)!} [(p-{\rm i}\alpha)\partial_p]^{i-k} [(q+{\rm i}\alpha)\partial_q]^{j-l} \frac{1}{p+q} (-\frac{p-{\rm i}\alpha}{q+{\rm i}\alpha})^n e^{(p+q)x -(p^2-q^2) {\rm i}t} \right|_{p=\zeta,q=\zeta^*},
\end{eqnarray}
and
\begin{eqnarray*}
&& \zeta=\zeta_r + {\rm i}\zeta_i,\ \
 \zeta_r= \pm\frac{\sqrt{3}}{12}\frac{K^2-4\alpha^2}{K},\ \ \zeta_i=\frac{1}{12}K + \frac{1}{3}\frac{\alpha^2}{K} + \frac{2}{3}\alpha,
\end{eqnarray*}
with
\begin{eqnarray*}
&& K= \left( 8\alpha^3 +108 + 12 \sqrt{12\alpha^3+81} \right)^{1/3}, \ \ \  (\alpha > -\frac{3}{2}2^{1/3}),
\end{eqnarray*}
where $a_k^{(\nu)}$ are complex constants and need to satisfy the relations:
\begin{eqnarray}\label{theoremeq04}
&& a^{(\nu+1)}_{k}=\sum^{k}_{j=0} \frac{2^{j+2} (p-{\rm i}\alpha)^2 + (-1)^{j}  \frac{2{\rm i}}{p-{\rm i}\alpha} +2{\rm i}\alpha(p-{\rm i}\alpha)}{ (j+2)!} a^{(\nu)}_{k-j},
\ \ \nu=0,1,2,\cdots.
\end{eqnarray}

\section{ Dynamics of rogue wave solutions}

In this section, we present the dynamics analysis of rogue wave solutions to the 1D YO system in detail.
To this end, we fix the parameter $\zeta_r= \frac{\sqrt{3}}{12}\frac{K^2-4\alpha^2}{K}$ without loss of generality.
Meanwhile, due to the fact that the long wave $L$ is a real-valued  function and its rogue wave structure is always of bright,
in what follows, we omit the discussion of the long wave component and only consider the dynamical properties of the complex short wave component $S$.

\subsection{Fundamental rogue wave}

According to \textbf{Theorem 2.3}, in order to obtain the first-order rogue wave, we need to take $N=1$ in Eqs.(\ref{theoremeq01})--(\ref{theoremeq04}). For simplicity, we set $a^{(0)}_0=b^{(0)}_0=1$, $a^{(0)}_1=b^{(0)}_1=0$,
then the functions $f$ and $g$ take the form
\begin{eqnarray}
&& f=\frac{1}{\zeta_r}e^{2\zeta_r(x+2\zeta_i t)} ( \theta \theta^* +\theta_0 ),\\
&& g=\frac{1}{\zeta_r}e^{2\zeta_r(x+2\zeta_i t)} \left[- \frac{ \zeta_r + {\rm i}(\zeta_i-\alpha) }{ \zeta_r - {\rm i}(\zeta_i-\alpha) } \right]
\left[ \left(\theta -\frac{1}{2} + \frac{1}{2}{\rm i}\right) \left(\theta^* +\frac{1}{2} + \frac{1}{2}{\rm i} \right) +\theta_0 \right],
\end{eqnarray}
with
\begin{eqnarray*}
&& \theta= (k_1 +{\rm i}k_2)x + (h_1 +{\rm i}h_2)t + (l_1 +{\rm i}l_2),\\
&& k_1=\frac{1}{2} (\alpha + \zeta_r - \zeta_i),\ \
k_2=\frac{1}{2} (\alpha - \zeta_r- \zeta_i),\\
&& h_1=\alpha(\zeta_i-\zeta_r)+\zeta^2_r + 2\zeta_r \zeta_i -\zeta^2_i,\ \
h_2=\alpha(\zeta_i+\zeta_r)+\zeta^2_r - 2\zeta_r \zeta_i -\zeta^2_i,\\
&& l_1= \frac{\zeta_i-\alpha}{4\zeta_r} - \frac{1}{4},\ \
l_2= \frac{\zeta_i-\alpha}{4\zeta_r} + \frac{1}{4}, \\
&& \theta_0=\frac{(\zeta_i-\alpha)^2}{8\zeta^2_r} + \frac{1}{8}.
\end{eqnarray*}
Thus, the fundamental rogue wave solution reads
\begin{eqnarray}
\label{rogue-sw-01}&& S = e^{-{\rm i} [\alpha x + (h+\alpha^2) t]  } \left[- \frac{ \zeta_r + {\rm i}(\zeta_i-\alpha) }{ \zeta_r - {\rm i}(\zeta_i-\alpha) } \right]
\left[ 1 + \frac{ {\rm i}(L_1+L_2) -1/2 }{ L^2_1 + L^2_2 +\theta_0 } \right],\\
\label{rogue-lw-02}&& L= h +  4 \frac{ (k_1L_1+k_2L_2)^2 - (k_2L_1-k_1L_2)^2 -(k^2_1+k^2_2)\theta_0 }{ (L^2_1 + L^2_2 +\theta_0)^2 },
\end{eqnarray}
where $L_1=k_1 x+ h_1 t +l_1$ and $L_2=k_2 x+ h_2 t +l_2$.

It is found that
the modular square of the short-wave component $|S|^2$ possesses critical points
\begin{eqnarray}
&& (x_1,t_1) = \left( \frac{1}{2\zeta_r}, 0 \right),\\
&& (x_2,t_2) = \left( \frac{1}{2}\frac{\mu_1(\alpha - 2\zeta_i)}{\zeta_r \Delta} + \frac{1}{2\zeta_r}, \frac{1}{4}\frac{\mu_1}{\zeta_r \Delta} \right),\\
&& (x_3,t_3) = \left( -\frac{1}{2}\frac{\mu_2(\alpha \zeta_i + \zeta^2_r - \zeta^2_i)}{\zeta^2_r \Delta} + \frac{1}{2\zeta_r}, \frac{1}{4}\frac{\mu_2(\alpha-\zeta_i)}{\zeta^2_r \Delta} \right),
\end{eqnarray}
with
\begin{equation*}
\mu_1=\pm\sqrt{3(\alpha-\zeta_i)^2-\zeta^2_r},\ \
\mu_2=\pm\sqrt{3\zeta^2_r-(\alpha-\zeta_i)^2},\ \
\Delta=(\alpha-\zeta_i)^2+\zeta^2_r.
\end{equation*}
Note that $(x_3, t_3)$ are also two characteristic points, at which the values of the amplitude are zero.

At these points, the local quadratic forms are
\begin{eqnarray}
\nonumber && H(\tilde{x},\tilde{t}) = \left.\left[ \frac{\partial^2 |S|^2}{\partial x^2} \frac{\partial^2 |S|^2}{\partial t^2}  - \left( \frac{\partial^2 |S|^2}{\partial x \partial t} \right)^2 \right]\right|_{(\tilde{x}, \tilde{t})},\\
&& H(x_1,t_1)=-\frac{16384 \zeta^{10}_r \mu^2_1 \mu^2_2  }{\Delta^4},\ \
   H(x_2,t_2)=\frac{64 \zeta^{10}_r \mu^2_1 \Delta^2 }{ (\alpha-\zeta_i)^{10} },\ \
   H(x_3,t_3)=64\mu^2_2 \Delta^2,
\end{eqnarray}
and the second derivatives are
\begin{eqnarray}
\nonumber && H_1(\tilde{x},\tilde{t}) = \left. \frac{\partial^2 |S|^2}{\partial x^2} \right|_{(\tilde{x}, \tilde{t})},\\
&& H_1(x_1,t_1)=\frac{192 \zeta^4_r [ (\alpha-\zeta_i)^2 - \zeta^2_r ] }{\Delta^2},\ \
   H_2(x_2,t_2)=-\frac{6 \zeta^4_r \Delta }{(\alpha-\zeta_i)^4},\ \
   H_3(x_3,t_3)= 6\Delta.
\end{eqnarray}

Based on above analysis, the fundamental rogue wave can be classified into three patterns:

(a)\ \ Dark state ($ -\frac{3}{2}2^{1/3} <\alpha \leqslant -\frac{2}{3}3^{2/3} $):
\ \  two local maximums at $(x_2,t_2)$ with the $|S|$'s amplitude $\frac{\sqrt{\Delta}}{\zeta_i-\alpha}$, one local minimum at $(x_1,t_1)$ with the $|S|$'s amplitude $-\frac{\mu^2_2}{\Delta}$.
Especially, when $\alpha = -\frac{2}{3}3^{2/3}$, the local minimum is located at the characteristic point $(x_1,t_1)=(x_3,t_3)=(\frac{1}{3}3^{5/6},0)$.

(b)\ \ Intermediate state ($ -\frac{2}{3}3^{2/3} < \alpha< 0$):
\ \ two local maximums at $(x_2,t_2)$ with the $|S|$'s amplitude $\frac{\sqrt{\Delta}}{\zeta_i-\alpha}$, two local minimums at two characteristic points $(x_3,t_3)$.

(c)\ \ Bright state ($ \alpha \geqslant 0$):
two local minimums at two characteristic points $(x_3,t_3)$, one local maximums at $(x_1,t_1)$ with the $|S|$'s amplitude $\frac{\mu^2_2}{\Delta}$. Particularly,
$\alpha =0$, the local maximum is located at $(x_1,t_1)=(x_2,t_2)=(\frac{1}{3}3^{1/2},0)$.

\begin{figure}[!htbp]
\centering
\includegraphics[width=4.77in,height=1.55in]{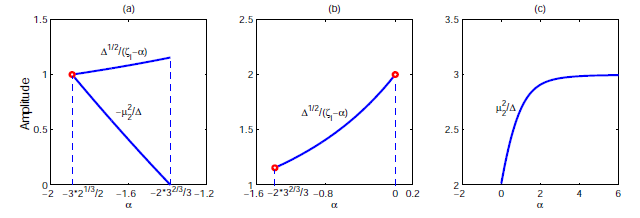}
\caption{The extreme values of $|S|$ with the parameter $\alpha$: (a) dark state; (b) intermediate state;  and (c) bright state.  \label{amplitude-fig}}
\end{figure}

\begin{figure}[!htbp]
\centering
{\includegraphics[height=1.6in,width=5.6in]{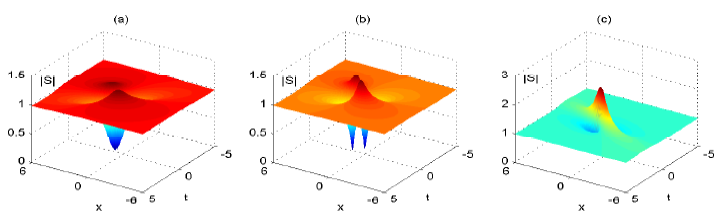}}
\caption{  First-order rogue wave of YO system: (a) dark state $\alpha=-\frac{7}{5}$; (b) intermediate state $\alpha=-1$;  and (c) bright state $\alpha=\frac{1}{3}$. \label{first-fig} }
\end{figure}


At the extreme points, the evolution of the amplitudes for the short wave with the parameter $\alpha$ is exhibited in Fig.\ref{amplitude-fig}.
It can be clearly seen that, for the dark state, as $\alpha$ changes from $-\frac{3}{2}2^{1/3}$  to $-\frac{2}{3}3^{2/3}$, the maximal amplitudes increase from $1$ to $\frac{2}{3}\sqrt{3}$ while the minimal one decreases from $1$ to $0$; for the intermediate state, as
$\alpha$ changes from $-\frac{2}{3}3^{2/3}$ to $0$,
the maximal amplitudes increase from $(\frac{2}{3}\sqrt{3}$ to $2$ while the minimal amplitude is always zero; for the bright state, as
$\alpha \ge 0$, the maximal amplitude is changes from $2$ to its asymptotic value of $3$ while the minimal value is always zero.

Fig.\ref{first-fig} displays three patterns of fundamental rogue wave for short wave component.
In three cases, the amplitudes of the short wave uniformly approach to the background $1$ as $(x, t)$ goes to infinity.
Fig.\ref{first-fig} (a) exhibits a dark rogue wave, in which it has one hole falling to $0.0247$  at $(0.8420,0)$ and two humps with the
height $1.1450$ at $(0.1531,0.4981)$ and $(1.5309,-0.4981)$.
For Fig.\ref{first-fig} (b), as an example of the intermediate state of rogue wave,
it attains its maximums $1.3118$ at $(0.1645,0.3356)$ and $(1.1780,-0.3356)$, and minimums $0$ at $(1.0975,0.2823)$ and $(0.2451,-0.2823)$.
In Fig.\ref{first-fig} (c), the amplitude of the short wave features a bright rogue wave, which possesses the two zero-amplitudes points $(1.0715,0.2168)$ and $(0.0966,-0.2168)$ and acquires a maximum of  $2.2965$ at $(0.5840,0)$.
This bright rogue wave is similar to the Peregrine soliton, but its structure possesses the moving zeroamplitudes
points and the varying peak height owing to the arbitrary parameter $\alpha$.

From Eqs.(\ref{rogue-sw-01})--(\ref{rogue-lw-02}), it is known that the family of first-order rogue solutions contains two free parameters $\alpha$ and $h$.
The latter one $h$ is merely a constant for defining the background of the long wave component.
Therefore, based on the previous discussion, it is found that the feature of rogue wave for the short wave component depends on
the parameter $\alpha$.
The choice of the parameter $\alpha$ determines these local waves¡¯ patterns, more specifically, the number, the position of extrema and zero point, and further the type and height of extrema.
We comment here that the same parameter is also introduced in the construction of dark-dark soliton solution for the coupled NLS system \cite{ohta2011general} and the coupled YO system \cite{chen2014multidark}, in which this treatment results in the generation of non-degenerate dark-dark soliton solution.
As interpreted in \cite{ohta2011general}, this parameter can be formally removed by the Galilean transformation in the scaler NLS equation, while the same copies cannot be removed simultaneously in the coupled NLS system.
For the YO system, it contains the long wave and short wave coupling and is not Galilean invariant,
so the introduction of the parameter $\alpha$ is necessary and essential for the construction of the general rogue wave solutions including intermediate and dark rogue wave ones.

\subsection{Higher-order rogue wave}
The second-order rogue wave solution is obtained from Eqs.(\ref{theoremeq01})--(\ref{theoremeq04}) with $N=2$.
In this case, setting $a^{(0)}_0=b^{(0)}_0=1$, $a^{(0)}_1=b^{(0)}_1=0$, $a^{(0)}_2=b^{(0)}_2=0$, we obtain
the functions $f$ and $g$ as follows
\begin{eqnarray}\label{second-solution}
f
=\begin{vmatrix}
m_{11}^{(1,1,0)} & m_{13}^{(1,0,0)}  \\
m_{31}^{(0,1,0)} & m_{33}^{(0,0,0)}
\end{vmatrix},\ \
g
=\begin{vmatrix}
m_{11}^{(1,1,1)} & m_{13}^{(1,0,1)}  \\
m_{31}^{(0,1,1)} & m_{33}^{(0,0,1)}
\end{vmatrix}\,,
\end{eqnarray}
where the elements are determined by
\begin{eqnarray*}
&& m_{11}^{(1,1,n)} = A^{(1)}_1 B^{(1)}_1
\left.\tilde{m}^{(n)} \right|_{p=\zeta,q=\zeta^*},\ \
 m_{13}^{(1,0,n)} = A^{(1)}_1 B^{(0)}_3
\left.\tilde{m}^{(n)} \right|_{p=\zeta,q=\zeta^*},\\
&& m_{31}^{(0,1,n)} = A^{(0)}_3 B^{(1)}_1
\left.\tilde{m}^{(n)} \right|_{p=\zeta,q=\zeta^*},\ \
 m_{33}^{(0,0,n)} = A^{(0)}_3 B^{(0)}_3
\left.\tilde{m}^{(n)} \right|_{p=\zeta,q=\zeta^*},
\end{eqnarray*}
with the differential operators
\begin{eqnarray*}
&& A^{(1)}_1 = a^{(1)}_0 (p- {\rm i}\alpha) \partial_p +a^{(1)}_1,\ \
B^{(1)}_1 = a^{(1)*}_0 (q+ {\rm i}\alpha) \partial_q +a^{(1)*}_1,\\
&& A^{(0)}_3 = \frac{1}{6} [(p- {\rm i}\alpha) \partial_p]^3 +a^{(0)}_3,\ \
B^{(0)}_3 = \frac{1}{6} [(q+ {\rm i}\alpha) \partial_q]^3 +a^{(0)*}_3,
\end{eqnarray*}
and
\begin{eqnarray*}
\tilde{m}^{(n)}=\frac{1}{p+q} \left(-\frac{p-{\rm i}\alpha}{q+{\rm i}\alpha}\right)^n e^{(p+q)x -(p^2-q^2) {\rm i}t}\,.
\end{eqnarray*}
In addition
\begin{eqnarray*}
&& a^{(1)}_{0}= 2 (p-{\rm i}\alpha)^2 +  \frac{{\rm i}}{p-{\rm i}\alpha} +{\rm i}\alpha(p-{\rm i}\alpha) ,\ \
 a^{(1)}_{1}= \frac{1}{3} \left[ 4 (p-{\rm i}\alpha)^2 -  \frac{{\rm i}}{p-{\rm i}\alpha} + {\rm i}\alpha(p-{\rm i}\alpha) \right].
\end{eqnarray*}

\begin{figure}[!htbp]
\centering
{\includegraphics[height=3.2in,width=5.6in]{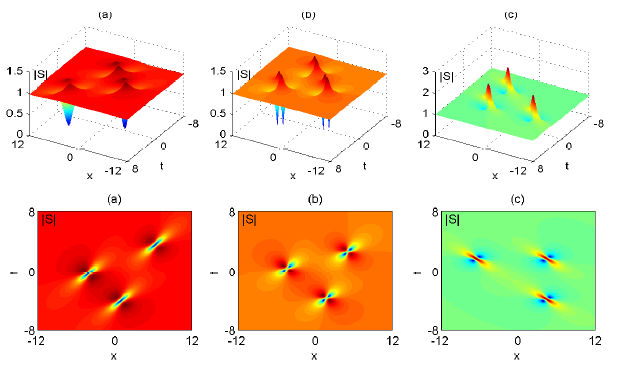}}
\caption{  Second-order rogue wave of YO system $a^{(0)}_3=85$: (a) dark state $\alpha=-\frac{7}{5}$; (b) intermediate state $\alpha=-1$;  and (c) bright state $\alpha=\frac{1}{3}$. \label{second-fig1} }
\end{figure}

\begin{figure}[!htbp]
\centering
{\includegraphics[height=3.2in,width=5.6in]{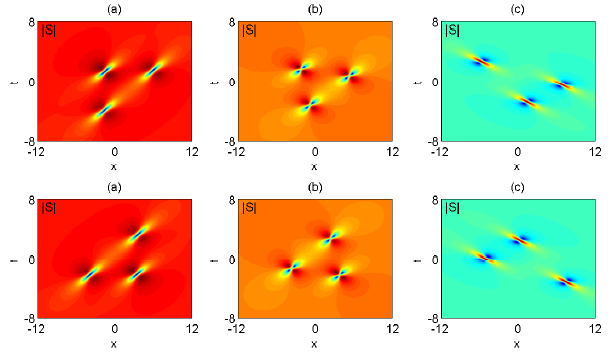}}
\caption{  Second-order rogue wave of YO system $a^{(0)}_3=50{\rm i}$(upper), $a^{(0)}_3=-50{\rm i}$(bottom): (a) dark state $\alpha=-\frac{7}{5}$; (b) intermediate state $\alpha=-1$;  and (c) bright state $\alpha=\frac{1}{3}$. \label{second-fig2} }
\end{figure}

Three groups of second-order rogue wave solutions with different values of the parameter $a^{(0)}_3$ are displayed in Fig.\ref{second-fig1}--\ref{second-fig2}.
As shown in these figures, the second-order rogue waves can be viewed as the superposition of three fundamental rogue waves, and
they have different dynamical behaviors for different values of the parameters $\alpha$ and $a^{(0)}_3$.
Here we first observer that when the values of the parameter $\alpha$ are still chosen same as the
figures displayed for the first-order rogue waves in which contain three elementary patterns, the structures of the
second-order rogue wave exhibit the triangle arrays of the fundamental dark, intermediate and bright rouge wave,
respectively.
Then it is easy to see that when the value of $\alpha$ is fixed, the dynamical behaviors for the second-order rogue waves depend on the values of $a^{(0)}_3$.
For example, two groups of rogue wave solutions with $a^{(0)}_3=50{\rm i}$ and $-50{\rm i}$ are shown in Fig.\ref{second-fig2},
the triangles for the arrays of elementary rogue waves are symmetric.

\begin{figure}[!htbp]
\centering
{\includegraphics[height=1.6in,width=5.6in]{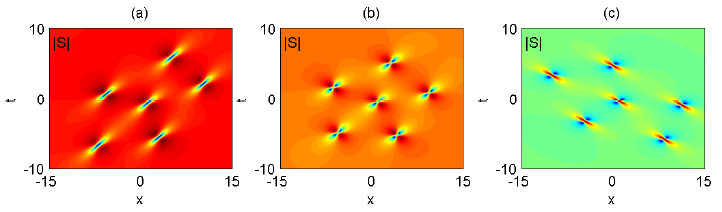}}
\caption{  Third-order rogue wave of YO system $a^{(0)}_5=2000$: (a) dark state $\alpha=-\frac{7}{5}$; (b) intermediate state $\alpha=-1$;  and (c) bright state $\alpha=0$.  \label{third-fig}}
\end{figure}

For the construction of third and higher -order rogue waves, which represent the superposition of more fundamental ones,
one need to take larger $N$ in (\ref{theoremeq01})--(\ref{theoremeq04}). The expressions is too complicated to illustrate here.
However we can show the dynamical structures of rogue waves graphically.
For $N=3$, we choose  $a^{(0)}_0=1$, $a^{(0)}_2=a^{(0)}_3=a^{(0)}_4=0$, $a^{(0)}_5=2000$ in Eqs.(\ref{theoremeq01})--(\ref{theoremeq04}), and plot the third-order rogue wave solution
in Fig.\ref{third-fig}. It can be seen that this third-order rogue waves exhibit  the superposition of six fundamental rogue waves and they constitute a shape of pentagon.

Finally, we would like to remark that whether the patterns of first-order rogue wave or the fundamental ones occur in the superposition for the higher-order case completely depends on the parameter $\alpha$ (see Fig.\ref{first-fig}--\ref{third-fig}).
In other words, three types of fundamental rogue waves and their higher-order superposition appear at three different intervals of $\alpha$, i.e., $(-\frac{3}{2}2^{1/3}, -\frac{2}{3}3^{2/3}]$ for dark state, $( -\frac{2}{3}3^{2/3}, 0)$ for intermediate state and $[0,+\infty)$ for bright state.
Therefore, there is no pattern of superposition among different types of fundamental rogue waves, for example, between bright ones and
dark ones.
Underling this fact is only a single copy of $\alpha$ is introduced in the 1D YO system.
In the construction of general solutions to the coupled YO system with multi-short wave components \cite{chen2015rational}, the multiple copies of $\alpha_i$ can be introduced which allows the superposition of different types of fundamental rogue waves by taking appropriate values of the parameters.
In addition, we comment that the 1D YO system with one short wave and one long wave coupling is different from the vector NLS equation representing two short wave coupling. As reported in \cite{zhao2016high,ling2016darboux}, two copies of $\alpha$ can be imposed and different fundamental rogue wave's superpositions can be exhibited.
The comparison reveals that the degree of freedom in the 1D YO system is less than one in the two-component NLS system.

\section{Summary and discussions}

In this paper, we derived general high-order rogue wave solutions for the 1D YO system by virtue of Hirota's  bilinear method.
These rogue wave solutions were obtained by the KP-hierarchy reduction technique and were expressed in terms of determinants
whose elements are algebraic formulae.
By choosing different values parameters in the rogue wave solutions,
we analytically and graphically studied the dynamics of first-, second- and third-order rogue wave solutions.
As a result, the fundamental (first-order) rogue waves are classified into three different patterns: bright, intermediate and dark states.
The higher-order rogue waves correspond to the superposition of fundamental rogue waves.
In particular, we should mention here that, in compared with the nonlinear Sch\"{o}dinger equation,
there exists an essential parameter $\alpha$ to control the pattern of rogue wave for both first- and higher-order rogue waves since the YO system does not possess the Galilean invariance.

Apart from the rogue wave appearing in the continuous models,
the rogue wave behaviors in discrete systems have recently drawn a lot of attention \cite{bludov2009rogue,ankiewicz2010discrete,ohta2014general}.
Paralleling to the novel patterns of rogue waves such as dark and intermediate ones in the continuous coupled systems with multiple waves,
the discrete counterparts of such rogue waves can be attained in discrete systems.
We have recently proposed an integrable semi-discrete analogue of the 1D YO system \cite{chen2016integrable}.
Thus the semi-discrete rogue wave, especially the semi-discrete dark and intermediate ones are worthy to be expected.
We will report the results on this topic in the future.

\section*{Acknowledgment}
Y.C. acknowledges support from the Global Change Research Program of
China (No. 2015CB953904), National Natural Science Foundation of China (Nos. 11675054, 11275072 and
11435005), and Shanghai Collaborative Innovation Center of Trustworthy Software for Internet of Things
(No. ZF1213).
B.F.F. is supported by National Natural Science Foundation of China (No.11428102).
M.K. is supported by JSPS Grant-in-Aid for Scientific Research (C-15K04909) and JST CREST.
Y.O. is partly supported by JSPS Grant-in-Aid for
Scientific Research (B-24340029, S-24224001, C-15K04909) and for
Challenging Exploratory Research (26610029).








\begin{thebibliography}{99}
%

\bibitem{kharif2009rogue}
C. Kharif, E. Pelinovsky, A. Slunyaev,
\newblock { Rogue Waves in the Ocean},
\newblock Springer, Berlin, 2009.

\bibitem{solli2007optical}
D. R.~Solli, C.~Ropers, P.~Koonath, B.~Jalali,
\newblock Optical rogue waves,
\newblock { Nature} 450 (2007) 1054--1057.

\bibitem{hohmann2010freak}
R.~H{\"o}hmann, U.~Kuhl, H.J. St{\"o}ckmann, L.~Kaplan, E.J.~Heller,
\newblock Freak waves in the linear regime: A microwave study,
\newblock { Phys. Rev. Lett.} 104 (2010) 093901.

\bibitem{montina2009non}
A.~Montina, U.~Bortolozzo, S.~Residori, F.T.~Arecchi,
\newblock Non-gaussian statistics and extreme waves in a nonlinear optical
  cavity,
\newblock { Phys. Rev. Lett.} 103 (2009) 173901.

\bibitem{bludov2009matter}
Y.V. Bludov, V.V.~Konotop, N.~Akhmediev,
\newblock Matter rogue waves,
\newblock { Phys. Rev. A} 80 (2009) 033610.

\bibitem{bludov2010vector}
Y.V. Bludov, V.V.~Konotop, N.~Akhmediev,
\newblock Vector rogue waves in binary mixtures of Bose-Einstein condensates,
\newblock {Eur. Phys. J. Special Topics} 185 (2010) 169--180.

\bibitem{ganshin2008observation}
A.N.~Ganshin, V.B.~Efimov, G.V.~Kolmakov, L.P.~Mezhov-Deglin, P.V.E. McClintock.
\newblock Observation of an inverse energy cascade in developed acoustic
  turbulence in superfluid helium,
\newblock { Phys. Rev. Lett.} 101 (2008) 065303.



\bibitem{stenflo2010rogue}
L. Stenflo, M. Marklund,
\newblock Rogue waves in the atmosphere,
\newblock { J. Plasma Phys.} 76 (2010) 293-295.

\bibitem{moslem2011langmuir}
W.M.~Moslem,
\newblock Langmuir rogue waves in electron-positron plasmas.
\newblock { Phys. Plasmas} 18 (2011) 032301.

\bibitem{bailung2011observation}
H.~Bailung, S.K.~Sharma, Y.~Nakamura,
\newblock Observation of peregrine solitons in a multicomponent plasma with
  negative ions,
\newblock { Phys. Rev. Lett.} 107 (2011) 255005.

\bibitem{shats2010capillary}
M.~Shats, H.~Punzmann, H.~Xia,
\newblock Capillary rogue waves,
\newblock {Phys. Rev. Lett.} 104 (2010) 104503.

\bibitem{yan2011vector}
Z.Y. Yan,
\newblock Vector financial rogue waves,
\newblock { Phys. Lett. A} 375 (2011) 4274--4279.


\bibitem{akhmediev2009waves}
N. Akhmediev, A. Ankiewicz, M. Taki,
\newblock Waves that appear from nowhere and disappear without a trace.
\newblock {Phys. Lett. A} 373 (2009) 675--678.

\bibitem{akhmediev2009extreme}
N. Akhmediev, J.M. Soto-Crespo, A. Ankiewicz,
\newblock Extreme waves that appear from nowhere: On the nature of rogue waves.
\newblock {Phys. Lett. A} 373 (2009) 2137--2145.


\bibitem{peregrine1983water}
D.H.~Peregrine,
\newblock Water waves, nonlinear Schr{\"o}dinger equations and their solutions,
\newblock { J. Aust. Math. Soc. B} 25 (1983) 16--43.

\bibitem{akhmediev2009rogue}
N. Akhmediev, A. Ankiewicz, J.M.~Soto-Crespo.
\newblock Rogue waves and rational solutions of the nonlinear Schr{\"o}dinger
  equation,
\newblock { Phys. Rev. E}  80 (2009) 026601.


\bibitem{kedziora2012second}
D.J. Kedziora, A. Ankiewicz, N. Akhmediev,
\newblock Second-order nonlinear Schr{\"o}dinger equation breather solutions in
  the degenerate and rogue wave limits,
\newblock {Phys. Rev. E}  85 (2012) 066601.

\bibitem{ankiewicz2011rogue}
A. Ankiewicz, D.J. Kedziora, N. Akhmediev,
\newblock Rogue wave triplets,
\newblock { Phys. Lett. A} 375 (2011) 2782--2785.

\bibitem{kedziora2011circular}
D.J. Kedziora, A. Ankiewicz, N. Akhmediev,
\newblock Circular rogue wave clusters,
\newblock { Phys. Rev. E} 84 (2011) 056611.

\bibitem{dubard2010multi}
P.~Dubard, P.~Gaillard, C.~Klein, V.B.~Matveev,
\newblock On multi-rogue wave solutions of the NLS equation and positon
  solutions of the KdV equation,
\newblock {Eur. Phys. J. Special Topics} 185 (2010) 247--258.

\bibitem{dubard2011multi}
P.~Dubard, V.B.~Matveev,
\newblock Multi-rogue waves solutions to the focusing NLS equation and the KP-I
  equation,
\newblock { Nat. Hazards Earth Syst. Sci.} 11 (2011) 667--672.

\bibitem{gaillard2011families}
P. Gaillard,
\newblock Families of quasi-rational solutions of the nls equation and
  multi-rogue waves.
\newblock { J. Phys. A: Math. Theor.} 44 (2011) 435204.

\bibitem{guo2012nonlinear}
B.L. Guo, L.M. Ling, Q.P.~Liu,
\newblock Nonlinear Schr{\"o}dinger equation: Generalized darboux
  transformation and rogue wave solutions.
\newblock { Phys. Rev. E} 85 (2012) 026607.

\bibitem{ohta2012general}
Y. Ohta, J.K. Yang,
\newblock General high-order rogue waves and their dynamics in the nonlinear
  Schr{\"o}dinger equation,
\newblock { Proc. R. Soc. London. Sect. A} 468 (2012) 1716--1740.

\bibitem{ankiewicz2010rogue}
A. Ankiewicz, J.M.~Soto-Crespo, N. Akhmediev,
\newblock Rogue waves and rational solutions of the Hirota equation,
\newblock {Phys. Rev. E}  81 (2010) 046602.

\bibitem{he2010generating}
J.S. He, H.R. Zhang, L.H. Wang, K. Porsezian, A. S. Fokas,
\newblock Generating mechanism for higher-order rogue waves.
\newblock {Phys. Rev. E} 87 (2013) 052914.


\bibitem{mu2016dynamic}
G. Mu, Z.Y. Qin,
\newblock Dynamic patterns of high-order rogue waves for Sasa--Satsuma
equation,
\newblock { Nonlin. Anal.: Real World Appl.} 31 (2016) 179--209.


\bibitem{ling2016multisoliton}
L.M. Ling, B.F. Feng, Z.N. Zhu,
\newblock Multi-soliton, multi-breather and higher order rogue wave solutions to the complex short pulse equation,
\newblock { Phys. D} 327 (2016) 13--29.


\bibitem{wang2017dynamics}
L. Wang, D.Y. Jiang, F.H. Qi, Y.Y. Shi, Y.C. Zhao,
\newblock Dynamics of the higher-order rogue waves for a generalized mixed nonlinear Schr{\"o}dinger model,
\newblock { Commun. Nonlinear Sci. Numer. Simulat.} 42 (2017) 502--519.


\bibitem{chen2015rational}
J.C. Chen, Y. Chen, B.F. Feng, K. Maruno,
\newblock Rational solutions to two-and one-dimensional multicomponent Yajima-Oikawa systems,
 {Phys. Lett. A} 379 (2015) 1510-1519.


\bibitem{bludov2009rogue}
Y.V. Bludov, V.V. Konotop, N. Akhmediev,
\newblock Rogue waves as spatial energy concentrators in arrays of nonlinear waveguides,
\newblock { Phys. Rev. E} 34 (2009) 3015-7.


\bibitem{ankiewicz2010discrete}
A. Ankiewicz, N. Akhmediev, J.M. Soto-Crespo,
\newblock Discrete rogue waves of the Ablowitz-Ladik and Hirota equations.
\newblock {Phys. Rev. E} 82 (2010) 026602.

\bibitem{ohta2014general}
Y. Ohta, J.K. Yang,
\newblock General rogue waves in the focusing and defocusing Ablowitz-Ladik equations,
\newblock { J. Phys. A: Math. Theor.} 47 (2014) 255201.


\bibitem{yan2010nonautonomous}
Z.Y. Yan,
\newblock Nonautonomous rogons in the inhomogeneous nonlinear Schr\"{o}dinger equation with variable coefficients,
\newblock { Phys. Lett. A} 374 (2010) 672-679.

\bibitem{yan2010threedimensional}
Z.Y. Yan,  V.V. Konotop, N. Akhmediev,
\newblock Three-dimensional rogue waves in nonstationary parabolic potentials,
\newblock { Phys. Rev. E} 82 (2010) 036610.

\bibitem{yan2010optical}
Z.Y. Yan,  C.Q. Dai,
\newblock Optical rogue waves in the generalized inhomogeneous higher-order nonlinear Schr\"{o}dinger equation with modulating coefficients,
\newblock { J. Opt.} 15 (2013) 064012.

\bibitem{wen2015generalized}
X.Y. Wen, Y.Q. Yang, Z.Y. Yan,
\newblock Generalized perturbation (n, M)-fold Darboux transformations and multi-rogue-wave structures for the modified self-steepening nonlinear Schr\"{o}dinger equation,
\newblock { Phys. Rev. E} 92 (2015) 012917.

\bibitem{yan2015twodimensional}
Z.Y. Yan,
\newblock Two-dimensional vector rogue wave excitations and controlling parameters in the two-component Gross-Pitaevskii equations with varying potentials,
\newblock { Nonlinear Dyn.} 79 (2015) 2515-2529.

\bibitem{yang2015rogue}
Y.Q. Yang, Z.Y. Yan, B.A. Malomed,
\newblock Rogue waves, rational solitons, and modulational instability in an integrable fifth-order nonlinear Schr\"{o}dinger equation,
\newblock {  Chaos} 25 (2015) 103112.

\bibitem{wen2016dynamics}
X.Y. Wen, Z.Y. Yan, Y.Q. Yang,
\newblock Dynamics of higher-order rational solitons for the nonlocal nonlinear Schr\"{o}dinger equation with the self-induced parity-time-symmetric potential,
\newblock {  Chaos} 26 (2016) 063123.

\bibitem{wen2017higherorder}
X.Y. Wen, Z.Y. Yan,
\newblock Higher-order rational solitons and rogue-like wave solutions of the (2+1)-dimensional nonlinear fluid mechanics equations,
\newblock {  Commun. Nonlinear Sci. Numer. Simulat.} 43 (2017) 311-329.




\bibitem{chabchoub2013observation}
A. Chabchoub, N. Akhmediev,
\newblock Observation of rogue wave triplets in water waves.
\newblock { Phys. Lett. A} 377 (2013) 2590--2593.

\bibitem{chabchoub2012observation}
A. Chabchoub, N. Hoffmann, M. Onorato, A. Slunyaev, A. Sergeeva, E. Pelinovsky, N. Akhmediev,
\newblock Observation of a hierarchy of up to fifth-order rogue waves in a water tank.
\newblock { Phys. Rev. E} 86 (2012) 056601.







\bibitem{ling2012highorder}
L.M. Ling, B.L. Guo, L.C. Zhao,
\newblock High-order rogue waves in vector nonlinear Schr{\"o}dinger equations,
\newblock {Phys. Rev. E} 89 (2014) 041201.

\bibitem{guo2011roguewave}
B.L. Guo, L.M. Ling,
\newblock Rogue wave, breathers and bright-dark-rogue solutions for the coupled Schr{\"o}dinger equations,
\newblock { Chin. Phys. Lett.} 28 (2011) 110202.


\bibitem{baronio2012solutions}
F. Baronio, A. Degasperis, M. Conforti, S. Wabnitz,
\newblock Solutions of the vector nonlinear Schr{\"o}dinger equations: evidence
  for deterministic rogue waves,
\newblock { Phys. Rev. Lett.} 109 (2012) 044102.


\bibitem{zhao2013roguewave}
L.C. Zhao, J. Liu,
\newblock Rogue-wave solutions of a three-component coupled nonlinear Schr{\"o}dinger equation,
\newblock { Phys. Rev. E} 87 (2013) 013201.



\bibitem{baronio2013solutions}
F. Baronio, M. Conforti, A. Degasperis, S. Lombardo,
\newblock Rogue waves emerging from the resonant interaction of three waves,
\newblock { Phys. Rev. Lett.} 111 (2013) 114101.



\bibitem{zhao2016high}
L.C. Zhao, B.L. Guo, L.M. Ling,
\newblock High-order rogue wave solutions for the coupled nonlinear Schr{\"o}dinger equations-II.
\newblock { J. Math. Phys. } 57 (2016) 043508.


\bibitem{ling2016darboux}
L.M. Ling, L.C. Zhao, B.L. Guo,
\newblock Darboux transformation and classification of solution for mixed coupled nonlinear Schr{\"o}dinger equations,
\newblock { Commun. Nonlinear Sci. Numer. Simulat.} 32 (2016) 285-304.

\bibitem{mu2015dynamics}
G. Mu, Z.Y. Qin, R. Grimshaw,
\newblock Dynamics of rogue waves on a multisoliton background in a vector nonlinear Schr{\"o}dinger equation,
\newblock { SIAM J. Appl. Math.} 75 (2015) 1-20.



\bibitem{zhai2013multirogue}
B.G. Zhai, W.G. Zhang, X.L. Wang, H.Q. Zhang,
\newblock Multi-rogue waves and rational solutions of the coupled nonlinear Schr{\"o}dinger equations,
\newblock {Nonlin. Anal.: Real World Appl.} 14 (2013) 14-27.


\bibitem{wang2015roguewave}
X. Wang, Y.Q. Li, F. Huang, Y. Chen,
\newblock Rogue wave solutions of AB system,
\newblock { Commun. Nonlinear Sci. Numer. Simulat.} 20 (2015) 434-442.


\bibitem{zhang2017solitons}
Y. Zhang, J.W. Yang, K.W. Chow, C.F. Wu,
\newblock Solitons, breathers and rogue waves for the coupled Fokas-Lenells
system via Darboux transformation,
\newblock { Nonlin. Anal.: Real World Appl.} 33 (2017) 237-252.

\bibitem{ohta2012rogue}
Y. Ohta, J.K. Yang,
\newblock Rogue waves in the Davey-Stewartson I equation,
\newblock { Phys. Rev. E} 86 (2012) 036604.

\bibitem{ohta2013dynamics}
Y. Ohta, J.K. Yang,
\newblock Dynamics of rogue waves in the Davey-Stewartson II equation,
\newblock {J. Phys. A: Math. Theor.} 46 (2013) 105202.

\bibitem{mu2017solitons}
G. Mu, Z.Y. Qin,
\newblock Two spatial dimensional N-rogue waves and their dynamics
in Mel'nikov equation,
\newblock {Nonlin. Anal.: Real World Appl.} 18 (2014) 1--13.


\bibitem{wen2015modulational}
X.Y. Wen, Z.Y. Yan
\newblock Modulational instability and higher-order rogue waves with parameters modulation in a coupled integrable AB system via the generalized Darboux transformation,
\newblock {Chaos} 25 (2015) 123115.

\bibitem{wen2016higherorder}
X.Y. Wen, Z.Y. Yan, B.A. Malomed,
\newblock Higher-order vector discrete rogue-wave states in the coupled Ablowitz-Ladik equations: Exact solutions and stability,
\newblock {Chaos} 26 (2016) 123110.



\bibitem{zakharov1972collapse}
V.E. Zakharov,
\newblock Collapse of langmuir waves,
 {Sov. Phys. JETP} 35 (1972) 908.


\bibitem{yajima1976formation}
N. Yajima, M. Oikawa,
\newblock Formation and interaction of sonic-langmuir solitons--inverse
  scattering method.
\newblock {Prog. Theor. Phys.} 56 (1976) 1719--1739.

\bibitem{zakharov1972collapse}
V.E. Zakharov,
\newblock Collapse of langmuir waves,
\newblock {Sov. Phys. JETP}  35 (1972) 908--914.



\bibitem{djordjevic1977two}
V.D.~Djordjevic, L.G.~Redekopp,
\newblock On two-dimensional packets of capillary-gravity waves.
\newblock {J. Fluid Mech.} 79 (1977) 703--714.

\bibitem{grimshaw1977modulation}
R.H.J. Grimshaw,
\newblock The modulation of an internal gravity-wave packet and the resonance
  with the mean motion,
\newblock {Stud. Appl. Math.} 56 (1977) 241--266.


\bibitem{ma1979some}
Y.C. Ma, L.G.~Redekopp,
\newblock Some solutions pertaining to the resonant interaction of long and
  short waves,
\newblock {Phys. Fluids} 22 (1979) 1872.

\bibitem{ma1978complete}
Y.C. Ma,
\newblock Complete solution of the long wave--short wave resonance equations,
{Stud. Appl. Math.} 59 (1978) 201.

\bibitem{kivshar1992stable}
Y.S. Kivshar,
\newblock Stable vector solitons composed of bright and dark pulses,
\newblock {Opt. Lett.} 17 (1992) 1322--1324.

\bibitem{chowdhury2008long}
A. Chowdhury, J.A. Tataronis,
\newblock Long wave--short wave resonance in nonlinear negative refractive
  index media,
\newblock {Phys. Rev. Lett.} 100 (2008) 153905.




\bibitem{cheng1992constraints}
Y. Cheng,
\newblock Constraints of the Kadomtsev-Petviashvili hierarchy,
\newblock {J. Math. Phys.} 33 (1992) 3774.


\bibitem{loris1997bilinear}
I. Loris, R. Willox,
\newblock Bilinear form and solutions of the $k$-constrained Kadomtsev-Petviashvili hierarchy,
\newblock {Inverse Problems} 13 (1997) 411.


\bibitem{wing2013rogue}
K.W.~Chow, H.N.~Chan, D.J.~Kedziora, R.H.J. Grimshaw,
\newblock Rogue wave modes for the long wave--short wave resonance model,
 {J. Phys. Soc. Jpn.} 82 (2013) 074001.

\bibitem{chen2014dark}
S.H. Chen, P. Grelu, J.M.~Soto-Crespo,
\newblock Dark-and bright-rogue-wave solutions for media with long-wave-short-wave resonance,
{Phys. Rev. E} 89 (2014) 011201.

\bibitem{chen2014darboux}
S.H. Chen,
\newblock Darboux transformation and dark rogue wave states arising from two-wave resonance interaction,
 {Phys. Lett. A} 378 (2014) 1095.



\bibitem{ohta2011general}
Y. Ohta, D.S. Wang, J.K. Yang,
\newblock General $N$-dark-dark solitons in the coupled nonlinear Schr{\"o}dinger equations,
\newblock {Stud. Appl. Math.} 127 (2011) 345--371.


\bibitem{chen2014multidark}
J.C. Chen, Y. Chen, B.F. Feng, K. Maruno,
\newblock Multi-dark soliton solutions of the two-dimensional multi-component Yajima-Oikawa systems,
 {J. Phys. Soc. Jpn.} 84 (2015) 034002.



\bibitem{chen2016integrable}
J.C. Chen, Y. Chen, B.F. Feng, K. Maruno, Y. Ohta,
\newblock An integrable semi-discretization of the coupled Yajima-Oikawa system,
 {J. Phys. A: Math. Theor.} 49 (2016) 165201.



\end{thebibliography}
\end{document}